\documentclass[fleqn,11pt,twoside]{article}
\usepackage{amsthm}
\usepackage{amsmath, amscd}%[fleqn,intlimits]
\usepackage{graphicx}
\usepackage{lscape}
\usepackage{amsmath}%[fleqn,intlimits]
\usepackage{graphicx}
\usepackage{latexsym}
\usepackage[all]{xy}
\makeatletter

\newcommand{\Name}[1]{\begin{flushleft}
                       \LARGE \bf #1
                       \end{flushleft}\vspace{-3mm}}

\newcommand{\Author}[1]{\begin{flushleft}
                       \it #1 \end{flushleft}}

\newcommand{\Address}[1]{\begin{flushleft}
                       \it #1 \end{flushleft}}

\newcommand{\FirstPageHead}[5]{
\begin{flushleft}
\raisebox{8mm}[0pt][0pt]
{\footnotesize \sf
\parbox{150mm}{ \qquad
 #1 #2 #3 
#4\hfill {\sc #5}}}\vspace{-13mm}
\end{flushleft}}

\newcommand{\evenhead}{Author \ name}
\newcommand{\oddhead}{Article \ name}

\renewcommand{\@evenhead}{
\hspace*{-3pt}\raisebox{-15pt}[\headheight][0pt]{\vbox{\hbox to \textwidth
{\thepage \hfil \evenhead}\vskip4pt \hrule}}}
\renewcommand{\@oddhead}{
\hspace*{-3pt}\raisebox{-15pt}[\headheight][0pt]{\vbox{\hbox to \textwidth
{\oddhead \hfil \thepage}\vskip4pt\hrule}}}
\renewcommand{\@evenfoot}{}
\renewcommand{\@oddfoot}{}

\long\def\@makecaption#1#2{%
  \vskip\abovecaptionskip
  \sbox\@tempboxa{\small \textbf{#1.}\ \ #2}%
  \ifdim \wd\@tempboxa >\hsize
    {\small \textbf{#1.}\ \ #2}\par
  \else
    \global \@minipagefalse
    \hb@xt@\hsize{\hfil\box\@tempboxa\hfil}%
  \fi
  \vskip\belowcaptionskip}

\newcommand{\JNMPnumberwithin}[3][\arabic]{%
  \@ifundefined{c@#2}{\@nocounterr{#2}}{%
    \@ifundefined{c@#3}{\@nocnterr{#3}}{%
      \@addtoreset{#2}{#3}%
      \@xp\xdef\csname the#2\endcsname{%
        \@xp\@nx\csname the#3\endcsname .\@nx#1{#2}}}}%
}

\newcommand{\resetfootnoterule} {
  \renewcommand\footnoterule{%
  \kern-3\p@
  \hrule\@width.4\columnwidth
  \kern2.6\p@}
}

\renewcommand{\footnoterule}{}

\newcommand{\be}{\begin{equation}}
\newcommand{\ee}{\end{equation}}
\newcommand{\ba}{\hspace*{-5pt}\begin{array}}
\newcommand{\ea}{\end{array}}
\newcommand{\p}{\partial}

\makeatother
\numberwithin{equation}{section}
\theoremstyle{definition}

\renewcommand{\ba}{\begin{array}}
\renewcommand{\ea}{\end{array}}
\newcommand{\beg}{\begin{eqnarray}}
\newcommand{\eeq}{\end{eqnarray}}
\newcommand{\bg}{\begin{eqnarray*}}

\newcommand{\ed}{\end{eqnarray*}}
\newcommand{\n}{\newline\hfill}

\renewcommand{\p}{\partial} 
 
\newcommand{\notlhd}{\lhd\kern-.8em{/}\ } 
\newcommand{\notexist}{\ \exists\kern-.5em{\raise.1em\hbox{/}}\ }

\newcommand{\pde}[2]{\frac{\p #1}{\p #2}}

\newcommand{\inp}{{\mbox{\vbox{\hrule width0ex\hbox{\vrule
 height0ex\kern3.8pt
\vbox{\kern2.5pt}\kern3.8pt \vrule height1.6ex}
\hrule width1.6ex}}}}

\begin{document}

%\Large

\renewcommand{\evenhead}{}
\renewcommand{\oddhead}{}

% Title

\thispagestyle{empty}

\begin{flushleft}
\footnotesize \sf
\end{flushleft}

\FirstPageHead{\ }{\ }{\ }
{ }{{
{ }}}
%\copyrightnote{2011}{M Euler and N Euler}

\Name{The Two-Component Camassa-Holm Equations CH(2,1) and CH(2,2):
  First-Order Integrating Factors and Conservation Laws}

\label{firstpage}

%\strut\hfill

%\strut\hfill

%\strut\hfill

\Author{Marianna Euler $^1$, Norbert Euler $^1$ and Thomas Wolf $^2$}

%\strut\hfill

%\strut\hfill

\Address{
%\noindent
$^1$ Department of Engineering Sciences and Mathematics\\ 
Lule\aa\ University of Technology\\
SE-971 87 Lule\aa, Sweden\\
Emails: Marianna.Euler@ltu.se; Norbert.Euler@ltu.se\\
$^2$ Department of Mathematics, Brock University\\
500 Glenridge Avenue, St.Catharines, Ontario, Canada L2S 3A1\\
twolf@brocku.ca
}

\vspace{1cm}

\noindent
{\bf Abstract:}\n
\noindent
Recently, Holm and Ivanov, proposed and studied a class of multi-component
generalisations of the Camassa-Holm equations [D D Holm and R I Ivanov,
Multi-component generalizations of the CH equation: geometrical 
aspects, peakons and numerical examples,
{\it J. Phys A: Math. Theor} {\bf 43}, 492001 (20pp), 2010]. We
consider two of those systems, denoted by Holm and Ivanov by CH(2,1) and
CH(2,2), and report a class of integrating factors and its
corresponding conservation laws for these two systems. In particular,
we obtain
the complete sent of first-order integrating factors for the systems
in Cauchy-Kovalevskaya form and evaluate the corresponding sets of
conservation laws for CH(2,1) and CH(2,2).

%\vspace{1cm}

%\strut\vfill

%\pagebreak

%%%%%%%%%%%%%%%%%%%%%%%%
\renewcommand{\theequation}{\arabic{section}.\arabic{equation}}
%\renewcommand{\theequation}{\arabic{equation}}

%%%%%%%%%%%%% TABLE OF CONTENTS %%%%%%%%%%%
%\tableofcontents

\section{Introduction}

It is well known that certain conservation laws of 
shallow water wave equations, such as the Camassa-Holm 
equation \cite{CamassaHolm}
and the the Degasperis-Procesi
equation \cite{Degasperis}, are useful to prove blow-up,
cf. the papers \cite{Constantin02}, \cite{Wu} and \cite{Wahlen}.
Furthermore, conservation laws play a central role
in the prove of the global existence (in time)
for solutions evolving from certain initial data, cf. the paper
\cite{ConstantinEscher}, and for proving the stability of peakons for both model
equations, cf. the papers \cite{ConstantinStrauss}, \cite{Lenells} 
and \cite{LinLiu}. In the context of the Camassa-Holm equation they 
are instrumental in the set-up of a theory of global weak solutions for
nonlinear nonlocal conservation laws, cf. the considerations in the 
papers \cite{Bressan-1}, \cite{Bressan-2} and \cite{Holden}

%\strut\hfill

In the current paper we derive all first-order integrating
factors and its corresponding conservation laws for some recently 
proposed multi-component generalizations of the 
Camassa-Holm equation \cite{Holm-Ivanov}.
We concentrate on two explicit systems, namely CH(2,1) and CH(2,2),
proposed by Holm and Ivanov in \cite{Holm-Ivanov} (see \ref{HI-1}) --
(\ref{HI-2}) and (\ref{CH22-1}) -- (\ref{CH22-2}) below).

%\strut\hfill

We recently reported in \cite{EE-CH-type} the complete set of 
first-order integrating factors and conservation laws
for a classs of Camassa-Holm type equations, which includes the
 Camassa-Holm 
equation \cite{CamassaHolm}
and the the Degasperis-Procesi
equation \cite{Degasperis}.
Our approach applied in this paper is based on the direct method 
described by Anco and Bluman
in their paper \cite{Anco-Bluman-1}, which can be applied to derive 
conservation laws of evolution equations that are in
Cauchy-Kovalevskaya form. We also refer the reader to
\cite{Wolf-1} and \cite{Wolf-2} for more details and alternate methods
for computing conservation laws for partial
differential equations and systems.

%\strut\hfill

Consider the 
two-component Camassa-Holm equations introduced and denoted by Holm and Ivanov
\cite{Holm-Ivanov} as CH(2,1), which has the following form:
\begin{subequations}
\begin{gather}
\label{HI-1}
\sigma_1 q_t+2q u_x+uq_x+\sigma \rho\rho_x=0\\[0.3cm]
\label{HI-2}
\rho_t+\rho u_x+u \rho_x=0,
\end{gather}
\end{subequations}
where
\begin{gather}
q=\sigma_1 u-u_{xx}+s
\end{gather}
and $s,\ \sigma$ and $\sigma_1$ are arbitrary constants. The
physically interesting cases are $\sigma=\pm 1$ and $\sigma_1=1$ or $\sigma_1=0$.
By defining the new dependent variables
\begin{subequations}
\begin{gather}
u:=U_1,\ u_x:=U_2\\[0.3cm] 
u_{xx}:=U_3,\ \rho:=U_4
\end{gather}
\end{subequations}
and the change of independent variables, 
\begin{gather}
\label{XT}
X:=t,\ T:=x, 
\end{gather}
we can write
system (\ref{HI-1}) -- (\ref{HI-2}) in the following
Cauchy-Kovalevskaya form:
\begin{subequations}
\begin{gather}
\label{4-system-a}
E_1:=U_{1,T}-U_2=0\\[0.3cm]
E_2:=U_{2,T}-U_3=0\\[0.3cm]
E_3:=U_{3,T}
-\sigma_1^2 U_1^{-1}U_{1,X}
+\sigma_1U_1^{-1}U_{3,X}
-3\sigma_1 U_2
+2U_1^{-1}U_2U_3
+\sigma U_1^{-2}U_4U_{4,X}\nonumber\\[0.3cm]
\qquad
+\sigma U_1^{-2}U_2U_4^2
-2s U_1^{-1}U_2=0\\[0.3cm]
\label{4-system-d}
E_4:=U_{4,T}+U_1^{-1}U_{4,X}+U_1^{-1}U_2U_4=0.
\end{gather}
\end{subequations}

%\strut\hfill

The second 2-component Camassa-Holm equation that we study in the
current paper, denoted by  CH(2,2),
has the form \cite{Holm-Ivanov}
\begin{subequations}
\begin{gather}
\label{CH22-1}
q_{1,t}+u_0 q_{1,x}+2q_1 u_{0,x}+u_1q_{2,x}+2q_2u_{1,x}=0\\[0.3cm]
\label{CH22-2}
q_{2,t}+u_0q_{2,x}+2q_2u_{0,x}=0,
\end{gather}
\end{subequations}
where
\begin{subequations}
\begin{gather}
q_1=u_1-u_{1,xx}+s_1\\
q_2=u_0-u_{0,xx}+3u_1^2-u_{1x}^2-2u_1u_{1,xx}
+4s_1u_1+s_2.
\end{gather}
\end{subequations}
Here $s_1,\ s_2$ are arbitrary constants.
By defining the new dependent variables
\begin{subequations}
\begin{gather}
u_0:=U_1,\ u_{0,x}:=U_2,\ u_{0,xx}:=U_3\\[0.3cm]
u_1:=U_4,\ u_{1,x}:=U_5,\ u_{1,xx}:=U_6
\end{gather}
\end{subequations}
and the change of independent variables (\ref{XT}),
we can present (\ref{CH22-1}) -- (\ref{CH22-2}) in the following
Cauchy-Kovalevskaya form:
\begin{subequations}
\begin{gather}
\label{CK-2-a}
E_1:=U_{1,T}-U_2=0\\[0.3cm]
E_2:=U_{2,T}-U_3=0\\[0.3cm]
E_3:=U_{3,T}
+12U_1^{-1}U_4^3U_5
-4U_1^{-1}U_4U_{4,X}
+2U_1^{-1}U_5U_{5,X}
-4s_1U_1^{-1}U_{4,X}\nonumber\\[0.3cm]
\qquad
+4U_5U_6
-4s_1U_5
+2U_1^{-1}U_2U_3
-6U_1^{-1}U_2U_4^2
+2U_1^{-1}U_2U_5^2
-2s_2U_1^{-1}U_2
\nonumber\\[0.3cm]
\qquad 
-4s_1U_1^{-1}U_2U_4
-12U_1^{-2}U_2U_4^4
+2U_1^{-1}U_6U_{4,X}
-8U_1^{-1}U_4^2U_5U_6\nonumber\\[0.3cm]
\qquad
+16s_1U_1^{-1}U_4^2U_{5}
+4U_1^{-2}U_4^2U_6U_{4,X}
-8s_1U_1^{-2}U_4^2U_{4,X}
+4U_1^{-2}U_4^2U_2U_3\nonumber\\[0.3cm]
\qquad 
+4U_1^{-2}U_4^2U_2U_5^2
+8U_1^{-2}U_2U_4^3U_6
-16s_1U_1^{-2}U_2U_4^3
-4s_2U_1^{-2}U_2U_4^2\nonumber\\[0.3cm]
\qquad 
+4U_1^{-2}U_4^2U_5U_{5,X}
-4U_1^{-1}U_3U_4U_5
+4s_2U_1^{-1}U_4U_5
-12U_1^{-2}U_4^3U_{4,X}\nonumber\\[0.3cm]
\qquad 
+2U_1^{-2}U_4^2U_{3,X}
+4U_1^{-2}U_4^3U_{6,X}
-4U_1^{-1}U_4U_5^3
-2U_1^{-2}U_4^2U_{1,X}\nonumber\\[0.3cm]
\qquad 
-U_1^{-1}U_{1,X}
+U_1^{-1}U_{3,X}
-3U_2=0\\[0.3cm]
E_4:=U_{4,T}-U_5=0\\[0.3cm]
E_5:=U_{5,T}-U_6=0\\[0.3cm]
E_6:=U_{6,T}
+4U_1^{-1}U_4U_5U_6
-8s_1U_1^{-1}U_4U_5
+2U_1^{-1}U_5^3
-3U_5
-U_1^{-1}U_{4,X}\nonumber\\[0.3cm]
\qquad
+U_1^{-1}U_{6,X}
-2U_1^{-2}U_4^2U_{6,X}
+6U_1^{-2}U_2U_4^3
-U_1^{-2}U_4U_{3,X}
+U_1^{-2}U_4U_{1,X}\nonumber\\[0.3cm]
\qquad
+6U_1^{-2}U_4^2U_{4,X}
-2U_1^{-2}U_4U_6U_{4,X}
+4s_1U_1^{-2}U_4U_{4,X}
-2U_1^{-2}U_2U_3U_4\nonumber\\[0.3cm]
\qquad
-2U_1^{-2}U_2U_4U_5^2
-4U_1^{-2}U_2U_4^2U_6
+8s_1U_1^{-2}U_2U_4^2
+2s_2U_1^{-2}U_2U_4\nonumber\\[0.3cm]
\qquad
-2U_1^{-2}U_4U_5U_{5,X}
+2U_1^{-1}U_3U_5
-2s_2U_1^{-1}U_5
+2U_1^{-1}U_2U_6\nonumber\\[0.3cm]
\label{CK-2-f}
\qquad
-2s_1U_1^{-1}U_2
-6U_1^{-1}U_4^2U_5=0.
\end{gather}
\end{subequations}

The above first-order Cauchy-Kovalevskaya systems can now be investigated for
integrating factors to derive conservation laws for the systems; 
which then leads to conservation
laws of the systems CH(1,1) and CH(2,2) in the original variables.

\strut\hfill

\section{General description}

In this section we breifly describe the direct method \cite{Anco-Bluman-1}
of integrating factors (or multipliers) for the general first-order 
Cauchy-Kovalevskaya system of six equations:
\begin{gather}
\label{our-gen-U1}
E_j:=U_{j,T}-F_j(U_1,\ldots,U_6,U_{1,X},\ldots,U_{6,X})=0,\quad
j=1,2,\ldots,6.
\end{gather}
Every conserved density, $\Phi^T$, and conserved flux, $\Phi^X$,
of system (\ref{our-gen-U1})
must satisfy
\begin{gather}
\left.
\vphantom{\frac{DA}{DB}}
D_T \Phi^T+D_X\Phi^X\right|_{\vec E=\vec 0}
=0,
\end{gather}
where, in general, both $\Phi^T$ and $\Phi^X$ are functions
of $X,T,U_j$ as well as $X$-derivatives of $U_j$.
Moreover, every $\Phi^T$ requires six integrating factors,
$\{\Lambda_1,\ \Lambda_2,\ldots,\Lambda_6\}$, which are directly related 
to the conserved density by the relation \cite{Anco-Bluman-1}
\begin{gather}
\Lambda_k=\hat E[U_k] \Phi^T,\qquad k=1,2,\ldots,6.
\end{gather}
Here $\hat E$ is the Euler Operator,
\begin{gather}
\hat E[U_k]:=
\pde{\ }{U_k}-D_T\circ \pde{\ }{U_{k,T}}
+\sum_{j=1}^{q}(-1)^{j}
D_X^j\circ \pde{\ }{U_{k,jX}},
\end{gather}
where we use the notation
\begin{gather*}
U_{k,jX}:=\frac{\p^j U_k}{\p X^j}.
\end{gather*}
The conditions on the integrating factors, 
$\{\Lambda_j\}$, of
system (\ref{our-gen-U1}) are 
\begin{gather}
\hat E[U_k]\left(\Lambda_1\, E_1+\Lambda_2\, E_2+\cdots+\Lambda_6\,
  E_6\right)=0, \qquad k=1,2,\ldots,6.
\end{gather}
However, since all integrating factors of system (\ref{our-gen-U1})
are adjoint symmetries of 
the system (\ref{our-gen-U1}), we can calculate  $\{\Lambda_j\}$
by the condition
\begin{gather}
\label{adj-symm-cond}
\left.
\left(
\ba{cccc}
L^*_{E_1}[U_1]&L^*_{E_2}[U_1]&\cdots
&L^*_{E_6}[U_1]\\[0.3cm]
L^*_{E_1}[U_2]&L^*_{E_2}[U_2]&\cdots
&L^*_{E_6}[U_2]\\[0.3cm]
\vdots&\vdots&\vdots&\vdots\\[0.3cm]
L^*_{E_1}[U_6]&L^*_{E_2}[U_6]&\cdots
&L^*_{E_6}[U_6]
\ea
\right)
\left(
\ba{c}
\Lambda_1\\[0.3cm]
\Lambda_2\\[0.3cm]
\vdots\\[0.3cm]
\Lambda_6
\ea\right)
\right|_{\vec E=\vec 0}
=\left(
\ba{c}
0\\[0.3cm]
0\\[0.3cm]
\vdots\\[0.3cm]
0
\ea\right)
\end{gather}
and then require the self-adjointness condition on 
$\{\Lambda_j\}$
(as integrating factors are variational quatities), namely
\begin{gather}
%\label{self-adjoint-cond}
\left(
\ba{llll}
L_{\Lambda_1}[U_1]&L_{\Lambda_1}[U_2]&\cdots&L_{\Lambda_1}[U_6]\\[0.3cm]
L_{\Lambda_2}[U_1]&L_{\Lambda_2}[U_2]&\cdots&L_{\Lambda_2}[U_6]\\[0.3cm]
\vdots            &\vdots            &\vdots&\vdots\\[0.3cm]
L_{\Lambda_6}[U_1]&L_{\Lambda_6}[U_2]&\cdots&L_{\Lambda_6}[U_6]
\ea
\right)
\left(
\ba{l}
E_1\\[0.3cm]
E_2\\[0.3cm]
\vdots\\[0.3cm]
E_6
\ea\right)\nonumber\\[0.3cm]
\label{self-adjoint-cond}
\qquad =\left(
\ba{llll}
L^*_{\Lambda_1}[U_1]&L^*_{\Lambda_2}[U_1]&\cdots&L^*_{\Lambda_6}[U_1]\\[0.3cm]
L^*_{\Lambda_1}[U_2]&L^*_{\Lambda_2}[U_2]&\cdots&L^*_{\Lambda_6}[U_2]\\[0.3cm]
\vdots              &\vdots              &\vdots&\vdots\\[0.3cm]
L^*_{\Lambda_1}[U_6]&L^*_{\Lambda_2}[U_6]&\cdots&L^*_{\Lambda_6}[U_6]
\ea
\right)
\left(
\ba{l}
E_1\\[0.3cm]
E_2\\[0.3cm]
\vdots\\[0.3cm]
E_6
\ea\right).\\[0.3cm]
%\label{self-adjoint-cond}
\end{gather}
Here $L$ is the linear operator and $L^*$ its adjoint:
\begin{subequations}
\begin{gather}
L_P[U_j]:=\pde{P}{U_j}+\sum_{i=1}^{p}\pde{P}{U_{j,iT}}\
D_T^i+\sum_{k=1}^{q}\pde{P}{U_{j,kX}}\ D_X^k\\[0.3cm]
L_P^*[U_j]:=\pde{P}{U_j}+\sum_{i=1}^p(-1)^i\ D_T^i\circ
\pde{P}{U_{j,iT}}+\sum_{k=1}^q(-1)^k\ D_X^k\circ \pde{P}{U_{j,kX}}.
\end{gather}
\end{subequations}
Note that the self-adjointness condition, (\ref{self-adjoint-cond}),
is independent of the form of the evolution system 
(\ref{our-gen-U1}) and only depends
on the functional arguments of $\{\Lambda_j\}$ as well as the number of
equations in the system.

\section{Integrating factors for system (\ref{4-system-a}) --
  (\ref{4-system-d}) and conservation laws for (\ref{HI-1}) --
(\ref{HI-2}):}
Solving conditions (\ref{adj-symm-cond}) and (\ref{self-adjoint-cond})
for system (\ref{4-system-a}) -- (\ref{4-system-d}),
the complete set of first-order integrating factors
$\{\Lambda_1,\ldots,\Lambda_4\}$, of the form
\begin{gather*}
\Lambda_j=\Lambda_j(X,T,U_1,\ldots,U_4,U_{1,X},\ldots,U_{4,X}),\quad
j=1,2,\ldots,4
\end{gather*}
for arbitrary $\sigma$, $\sigma_1\neq 0$ and $s$ is as follows:
\begin{subequations}
\begin{gather}
\Lambda_1=\lambda_1
U_4
+2\lambda_2\left(
s+\frac{3}{2}\sigma_1U_1-\frac{1}{2}U_3\right)\nonumber\\
-\lambda_3\left(\sigma_1U_{2,X}-2sU_1-\sigma U_4^2-3\sigma_1 U_1^2
+2U_1U_3\right)\sigma_1^{-1}\\[0.3cm]
\Lambda_2=-\lambda_2U_2
+\lambda_3 U_{1,X}\\[0.3cm]
\Lambda_3=
-\lambda_2 U_1
-\lambda_3 \sigma_1^{-1}U_1^2\\[0.3cm]
\Lambda_4=
\lambda_1U_1
+\lambda_2\sigma U_4
+2\lambda_3\sigma\sigma_1^{-1}U_1U_4
\end{gather}
\end{subequations}
where $\lambda_j$ are arbitrary constants. This leads to the following
three sets of conserved density, $\Phi^t$, and conserved flux, $\Phi^x$, for the original system (\ref{HI-1}) --
(\ref{HI-2}) (separated by means of the arbitrary $\lambda_1$,
$\lambda_2$ and $\lambda_3$, respectively ):
\begin{subequations}
\begin{gather} 
\Phi_1^t=\rho\\[0.3cm]
\Phi_1^x=u\rho
\end{gather}
\end{subequations}

\begin{subequations}
\begin{gather} 
\Phi_2^t=\sigma_1^2u-\sigma_1 u_{xx}
\\[0.3cm]
\Phi_2^x=
2su+\frac{1}{2}\sigma \rho^2
+\frac{3}{2}\sigma_1 u^2
-uu_{xx}
-\frac{1}{2}u_x^2
\end{gather}
\end{subequations}

\begin{subequations}
\begin{gather} 
\Phi_3^t=
\left(-\sigma_1 u_x^2
+\frac{1}{2}\sigma \rho^2
+\frac{1}{2}\sigma_1^2u^2
-\sigma_1 uu_{xx}+\frac{1}{2}\sigma_1 u_x^2\right)\sigma_1^{-1}
\\[0.3cm]
\Phi_3^x= 
\left(
\sigma_1 u_xu_t+su_x^2+\sigma u\rho^2
+\sigma_1 u^3
-u^2 u_{xx}\right)\sigma_1^{-1}
\end{gather}
\end{subequations}

Some special must be considered:

\strut\hfill

\noindent
{\bf Special Case 1:} $\sigma=0$ with $\sigma_1$ arbitrary, but
nonzero, and $s$ arbitrary. The integrating factors are as follows:
\begin{gather}
\Lambda_1=\frac{2U_3-3\sigma_1 U_1-2s}{(\sigma_1
  U_1-U_3+s)^{1/2}},\quad
\Lambda_2=0\\[0.3cm]
\Lambda_3=\frac{U_1}{(\sigma_1
  U_1-U_3+s)^{1/2}},\quad
\Lambda_4=0.
\end{gather}
and the corresponding conserved current for 
system (\ref{4-system-a}) -- (\ref{4-system-d}) is
\begin{subequations}
\begin{gather}
\Phi^t=\sigma_1 \left(\sigma_1 u-u_{xx}+s\right)^{1/2}\\[0.3cm]
\Phi^x=\left(\sigma_1 u-u_{xx}+s\right)^{1/2}.
\end{gather}
\end{subequations}
{\bf Special Case 2:} $\sigma=0$ with $\sigma_1=1$ 
and $s$ arbitrary. The integrating factors are as follows:
\begin{subequations}
\begin{gather}
\Lambda_1=-\frac{U_1W^3  H'(W)}{U_4}+2U_4H(W),\quad
\Lambda_2=0\\[0.3cm]
\Lambda_3=\frac{U_1W^3H'(W)}{U_4},\quad
\Lambda_4=2U_1\left(W H'(W)+H(W)\right),
\end{gather}
\end{subequations}
where $H(W)$ is an arbitrary differentiable function with
\begin{gather}
W:=\frac{U_4}{(U_1-U_3+s)^{1/2}}.
\end{gather}
The conserved current for system (\ref{4-system-a}) --
(\ref{4-system-d})
is then
\begin{subequations}
\begin{gather}
\Phi^t=H(w)\rho\\[0.3cm]
\Phi^x=H(w)u\rho.
\end{gather}
\end{subequations}
Here the argument, $w$, in the arbitrary function $H$, is
\begin{gather}
w:=\frac{\rho}{(u-u_{xx}+s)^{1/2}}.
\end{gather}
{\bf Special Case 3:} $\sigma$ arbitrary, but nonzero, 
with $\sigma_1=1$ and $s$ arbitrary. The integrating factors are
as follows:
\begin{subequations}
\begin{gather}
\Lambda_1=\frac{U_3-2U_1-s}{\sigma U_4},\quad
\Lambda_2=0,\\[0.3cm]
\Lambda_3=\frac{U_1}{\sigma U_4},\quad
\Lambda_4=\frac{sU_1-\sigma U_4^2+U_1^2-U_1U_3}{\sigma U_4^2}.
\end{gather}
\end{subequations}
The conserved current for system (\ref{4-system-a}) --
(\ref{4-system-d})
is then
\begin{subequations}
\begin{gather}
\Phi^t=\frac{u_{xx}-u-s}{\sigma\rho}\\[0.3cm]
\Phi^x=\frac{uu_{xx}-u^2-\sigma \rho^2-su}{\sigma \rho}.
\end{gather}
\end{subequations}
{\bf Special Case 4:}  $\sigma_1=0$
with $\sigma$ and $s$ arbitrary. The integrating factors are
as follows:
\begin{subequations}
\begin{gather}
\Lambda_1=(U_3-2s) H(X, W),\quad
\Lambda_2=U_2 H(X,W)\\[0.3cm]
\Lambda_3=U_1 H(X,W),\quad
\Lambda_4=-\sigma U_4 H(X,W),
\end{gather}
\end{subequations}
where $H$ is an arbitrary differentiable function and
\begin{gather}
W:=\frac{1}{2}
\left(-4sU_1-\sigma U_4^2+2U_1U_3+U_2^2\right).
\end{gather}
The conserved current for system (\ref{4-system-a}) --
(\ref{4-system-d})
is then
\begin{subequations}
\begin{gather}
\Phi^t=0\\[0.3cm]
\Phi^x=w H(t,w),
\end{gather}
\end{subequations}
where
\begin{gather}
w:=\frac{1}{2}
\left(-4su-\sigma \rho^2+2uu_{xx}+u_x^2\right).
\end{gather}
{\bf Special Case 5:}  $\sigma_1=0$ with
$\sigma$ arbitrary, but nonzero, 
and $s$ arbitrary. The integrating factors are
as follows:
\begin{subequations}
\begin{gather}
\Lambda_1=
\frac{2}{\sigma}\left(
sU_1+\frac{\sigma}{2}U_4^2-U_1U_3\right),\quad
\Lambda_2=0\\[0.3cm]
\Lambda_3=-\frac{U_1^2}{\sigma},\quad
\Lambda_4=2U_1U_4.
\end{gather}
\end{subequations}
The conserved current for system (\ref{4-system-a}) --
(\ref{4-system-d})
is then
\begin{subequations}
\begin{gather}
\Phi^t=\frac{\rho^2}{2}\\[0.3cm]
\Phi^x=\frac{1}{\sigma}\left(
su^2+\sigma u\rho^2-u^2u_{xx}\right).
\end{gather}
\end{subequations}

\section{Integrating factors for system (\ref{CK-2-a}) --
  (\ref{CK-2-f}) and conservation laws for (\ref{CH22-1}) --
(\ref{CH22-2}):}
Solving conditions (\ref{adj-symm-cond}) and (\ref{self-adjoint-cond})
for system (\ref{CK-2-a}) -- (\ref{CK-2-f}),
the complete set of first-order integrating factors
$\{\Lambda_1,\ldots,\Lambda_6\}$, of the form
\begin{gather*}
\Lambda_j=\Lambda_j(X,T,U_1,\ldots,U_6,U_{1,X},\ldots,U_{6,X}),\quad
j=1,2,\ldots,6
\end{gather*}
are the following:
\begin{subequations}
\begin{gather}
\Lambda_1=\lambda_1\left(
2U_6U_1+2U_3U_4+2U_4U_5^2
+4U_4^2U_6-6U_4^3
-2s_1U_1
-8s_1U_4^2-6U_1U_4
\right.\nonumber\\[0.3cm]
\qquad
\left.
-2s_2U_4+U_{5,X}\right)
+\lambda_2\left(
U_3+2U_5^2-4s_1U_4-3U_1-2s_2\right)
\nonumber\\[0.3cm]
\qquad
+\lambda_3\left(U_6-3U_4-2s_1\right)
+\lambda_4\left(
-8s_1^2U_4
-2s_1s_2
-s_1U_1
+2s_1U_3
-22s_1U_4^2\right.\nonumber
\\[0.3cm]
\left.
+12s_1U_4U_6
+2s_1U_5^2
-4s_2U_4
+2s_2U_6
-3U_1U_4
+U_1U_6
+4U_3U_4-2U_3U_6
-12U_4^3\right.\nonumber
\\[0.3cm]
\left.
+14U_4^2U_6
+4U_4U_5^2
-4U_4U_6^2
-2U_5U_6\right)Z^{-3/2}\nonumber\\[0.3cm]
+\frac{\lambda_5}{2}\left(
-8s_1U_4-2s_2-3U_1+2U_3-6U_4^2+4U_4U_6+2U_5^2\right)Z^{-1/2}\\[0.3cm]
\Lambda_2=-\lambda_1 U_{4,X}+\lambda_2 U_2+\lambda_3 U_5\\[0.3cm]
\Lambda_3=
2\lambda_1U_1U_4+
\lambda_2\left(U_1-2U_4^2\right)
+\lambda_3 U_4\nonumber\\[0.3cm]
+\lambda_4\left(-s_1U_1+8s_1U_4^2
+2s_2U_4
+U_1U_4
+U_1U_6
-2U_3U_4
+6U_4^3\right.\nonumber\\[0.3cm]
\left.-4U_4^2U_6
-2U_4U_5^2\right)Z^{-3/2}
+\frac{\lambda_5}{2}U_1Z^{-1/2}\\[0.3cm]
\Lambda_4=
\lambda_1\left(
2U_1U_5^2
+2U_1U_3
+2U_4U_{5,X}
-2s_2U_1
-3U_1^2-18U_1U_4^2
+U_{2,X}
\right.
\nonumber\\[0.3cm]
\qquad
\left.
-16s_1U_1U_4
+8U_1U_4U_6\right)
+\lambda_2\left(
24U_4^3-4U_3U_4-4U_4U_5^2-12U_4^2U_6
-2U_{5,X}
\right.
\nonumber\\[0.3cm]
\qquad
\left.
+24s_1U_4^2
-4s_1U_1+4s_2U_4\right)
+\lambda_3
\left(
U_3+4U_4U_6-3U_1
+2U_5^2-12U_4^2
\right.
\nonumber\\[0.3cm]
\qquad
\left.
-12s_1U_4
-2s_2\right)
%%%%%%%%%%%%%%%%%%%%%%%%%%%%%%%%%%%%%%%%%%%%
+2\lambda_4\left(
2s_1^2U_1
-48s_1^2U_4^2
-20s_1s_2U_4
-19s_1U_1U_4
-3s_1U_1U_6\right.\nonumber\\[0.3cm]
\left.
+20s_1U_3U_4
-84s_1U_4^3
+48s_1U_4^2U_6
+20s_1U_4U_5^2
-2s_2^2
-5s_2U_1
+4s_2U_3
-18s_2U_4^2\right.\nonumber\\[0.3cm]
\left.
+10s_2U_4U_6
+4s_2U_5^2
-3U_1^2
+5U_1U_3
-18U_1U_4^2
+8U_1U_4U_6
+5U_1U_5^2
+U_1U_6^2
-2U_3^2\right.\nonumber\\[0.3cm]
\left.
+18U_3U_4^2
-10U_3U_4U_6
-4U_3U_5^2
-36U_4^4
+42U_4^3U_6
+18U_4^2U_5^2
-12U_4^2U_6^2\right.\nonumber\\[0.3cm]
\left.
-10U_4U_5^2U_6
-2U_5^4\right)Z^{-3/2}
+\lambda_5U_1\left(
-2s_1+3U_4+U_6\right)Z^{-1/2}
\\[0.3cm]
%%%%%%%%%%%%%%%%%%%%%%%%%%%%%%%%%%%%%%%%%%%%%%%
\Lambda_5=\lambda_1\left(
4U_1U_4U_5-U_{1,X}-2U_4U_{4,X}\right)
+\lambda_2\left(4U_1U_5-4U_4^2U_5+2U_{4,X}\right)
\nonumber\\[0.3cm]
\qquad
+\lambda_3\left(U_2+4U_4U_5\right)
-2\lambda_4U_5\left(
s_1U_1
-8s_1U_4^2
-2s_2U_4
-U_1U_4
-U_1U_6
+2U_3U_4\right.\nonumber\\[0.3cm]
\left.
-6U_4^3
+4U_4^2U_6
+2U_4U_5^2\right)Z^{-3/2}
+\lambda_5 U_1U_5Z^{-1/2}\\[0.3cm]
%%%%%%%%%%%%%%%%%%%%%%%%%%%%%%%%%%%%%%%%%%%%%%
\Lambda_6=\lambda_1\left(
U_1^2+4U_1U_4^2\right)
-4\lambda_2 U_4^3
+\lambda_3\left(
U_1+2U_4^2\right)
+2\lambda_4\left(
3s_1U_1U_4
+8s_1U_4^3
+s_2U_1\right.\nonumber\\[0.3cm]
\left.+2s_2U_4^2
-U_1^2
-U_1U_3
+4U_1U_4^2
-U_1U_4U_6
-U_1U_5^2
-2U_3U_4^2
+6U_4^4\right.\nonumber\\[0.3cm]
\left.
-4U_4^3U_6
-2U_4^2U_5^2\right)Z^{-3/2}
+\lambda_5 U_1U_4Z^{-1/2},
\end{gather}
\end{subequations}
where
\begin{gather}
Z:=s_1U_4-s_2-U_1+U_3-3U_4^2+2U_4U_6+U_5^2.
\end{gather}
This leads to the following set of three conserved densities and conserved
flux for
the system (\ref{CH22-1}) -- (\ref{CH22-2}):
\begin{subequations}
\begin{gather}
\Phi_1^t=u_1u_{0,xx}+u_1^2u_{1,xx}-u_0u_1-2s_1u_1^2-2u_1^3\\[0.3cm]
\Phi_1^x=\left(u_0+u_1^2\right) u_{1,xt}
+2u_0u_1u_{0,xx}
+2u_0u_1u_{1,x}^2
+\left(4u_0u_1^2+u_0^2\right)u_{1,xx}\nonumber\\[0.3cm]
\qquad
-\frac{1}{2}u_0^2\left(6u_1+2s_1\right)
-u_0\left(6u_1^3+2s_2u_1+8s_1u_1^2\right)
-u_{0,x}u_{1,t}
\end{gather}
\end{subequations}

\begin{subequations}
\begin{gather}
\Phi_2^t=
2u_1u_{1,xx}
+u_{0,xx}
-u_0
-2u_1^2
+2u_{1,x}^2
-4s_1u_1\\[0.3cm]
\Phi_2^x=
-2u_1u_{1,xt}
+(u_0-2u_1^2)u_{0,xx}
-4u_1^3u_{1,xx}
+\frac{1}{2}u_{0,x}^2
+2(u_0-u_1^2)u_{1,x}^2\nonumber\\[0.3cm]
\qquad
-2(s_2+2s_1u_1)u_0
-\frac{3}{2}u_0^2
+2u_1^2(s_2+4s_1u_1+3u_1^2)
\end{gather}
\end{subequations}

\begin{subequations}
\begin{gather}
\Phi_3^t=u_{1,xx}-u_1
\\[0.3cm]
\Phi_3^x=
(u_0+2u_1^2)u_{1,xx}
+u_1u_{0,xx}
+u_{0,x}u_{1,x}
+2u_1u_{1,x}^2
-(2s_1+3u_1)u_0
\nonumber\\[0.3cm]
\qquad
-2u_1(s_2+3s_1u_1+2u_1^2).
\end{gather}
\end{subequations}

\begin{subequations}
\begin{gather}
\Phi_4^t=
2\left(s_1-u_{1,xx}+u_1\right)z^{-1/2}
\\[0.3cm]
\Phi_4^x=
2\left(
s_1u_0
+8s_1u_1^2
+2s_2u_1
+3u_0u_1
-u_0u_{1,xx}
-2u_{0,xx}u_1
+6u_1^3\right.\nonumber\\[0.3cm]
\left.
-4u_1^2u_{1,xx}
-2u_1u_{1,x}^2
\right)z^{-1/2}
\end{gather}
\end{subequations}

\begin{subequations}
\begin{gather}
\Phi_5^t=z^{1/2}\\[0.3cm]
\Phi_5^x=u_0z^{1/2},
\end{gather}
\end{subequations}
where
\begin{gather}
z:=s_1u_1-s_2-u_0+u_{0,xx}-3u_1^2+2u_1u_{1,xx}+u_{1,x}^2.
\end{gather}

\section{Concluding remarks}
We have derived the complete set of first-order integrating factors
for the systems CH(2,1) and CH(2,2) in Cauchy-Kovalevskaya form.
The corresponding sets of conservation laws related to these 
integrating factors have been derived for both these systems.
It would certainly be interesting to calculate higher-order
integrating factors, although the computations involved for such
calculations appear to be rather challenging. We aim to report 
some results in a future paper. 

We expect that the same method than was applied here could also be used 
to find conservation laws for more general CH-systems proposed in 
\cite{Holm-Ivanov} and \cite{Novikov}.

\section*{Acknowledgements}
ME and NE thanks the New Jersey Institute of Technology for there
hospitality during their sabbatical leave at this institute. 
ME and NE also thank the
Wenner-Gran Foundation and Lule\aa\ University of Technology
for financial support.

\end{document}